\documentclass{PoS}

\title{$2D$ Gauge  Field Theory}

\ShortTitle{$2D$ Gauge  Field Theory}

\author{\speaker{Andrey V. Koshelkin}\\
        \\ Moscow Institute for Physics and Engineering,
Kashirskoye sh., 31, 115409 Moscow, Russia
        E-mail: \email{$A\underline\
Kosh@internets.ru$}}

\author{Cheuk-Yin Wong\\ Physics Division, Oak Ridge National Laboratory, Oak
Ridge, TN 37831\\
        E-mail: \email{wongc@ornl.gov}}

\abstract{ We show from the action integral that under the assumption
  of longitudinal dominance and transverse confinement, QCD4 in (3+1)
  dimensional space-time can be approximately compactified into QCD2
  in (1+1) dimensional space-time.  In such a process, we find the
  relation between the coupling constant $g(2D)$ in QCD2 and the
  coupling constant $g(4D)$ in QCD4.  We also show that quarks and gluons
  in QCD2 acquire masses as a result of the compactification. }

\FullConference{Sixth International Conference on Quarks and Nuclear Physics,\\
        April 16-20, 2012\\
        Ecole Polytechnique, Palaiseau, Paris}

\begin{document}

\def\bb #1{{\bf #1}}

\section{Introduction}

Previously, t'Hooft showed that in the limit of large $N_c$ with
fixed $g^2N_c$ in single-flavor quantum chromodynamics in (3+1)
dimensional space-time (QCD4), planar diagrams with quarks at the
edges dominate, whereas diagrams with the topology of a fermion
loop or a wormhole are associated with suppressing factors of
$1/N_c$ and $1/N_c^2$, respectively \cite{tho74a}.  In this case a
simple-minded perturbation expansion with respect to the coupling
constant $g$ cannot describe the spectrum, while the $1/N_c$
expansion may be a reasonable concept, in spite of the fact that
$N_c$ is equal to 3 and is not very big. The dominance of the
planar diagrams allows one to consider QCD in one space and one
time dimensions (QCD2) and the physics resembles those of the dual
string or a flux tube, with the physical spectrum of a straight
Regge trajectory \cite{tho74b}. The properties of QCD in
two-dimensional space-time have been investigated by many workers
\cite{tho74a,tho74b,Fri93}. The flux tube picture of longitudinal
dynamics manifests itself in various aspects of hadron
spectroscopy \cite{Isg85}.

In the high-energy arena, the flux tube picture finds phenomenological
applications in hadron collisions and high-energy $e^-e^+$
annihilations \cite{And83,Won91,Won94,Won09}.  In these high-energy
processes, the (average) transverse hadron momenta of produced hadrons
are observed to be limited, as appropriate for particles confined in a
flux tube.  The idealization of the three-dimensional flux tube as a
one-dimensional string leads to the string fragmentation picture of
particle production in (1+1) space-time dimensions.  The classical
Lund model of string fragmentation has been quite successful in
describing particle production processes in hadron collisions and
$e^-e^+$ annihilations at high energies \cite{And83,Won94}.

With the successes of lower-dimensional descriptions of
high-energy collision processes in QCD, we would like to examine
the circumstances in high-energy processes under which QCD4 in
(3+1) dimensions can be compactified into QCD2 in (1+1)
dimensions, if one starts with the QCD4 action integral.  In such
a process, we will be able to find out how quantities in the
compactified QCD2 can be related to quantities in QCD4.  The
success of the compactification program will facilitate the
examination of some problems in QCD4 in the simpler dynamics of
QCD2.

\section{$4D \to 2D$   Compactification in the Action Integral }

The $4D$-action ${\cal A } (4D) $ resides in (3+1) dimensional
space-time.  There are however environments which allow the
compactification of the $4D$ action to reside in two-dimensional (1+1)
space-time, within which the dynamics can be greatly simplified.

We note that in hadron collisions and $e^-e^+$ annihilations at
high energies, the string fragmentation process occurs when a
valence quark pulls apart from a valence antiquark longitudinally
at high energies. It is therefore reasonable to conceive that the
QCD4 compactification can take place under the dominance of
longitudinal dynamics, not only of the leading valence quark and
antiquark pair, but also the produced $q\bar q$ parton pairs.  In
the Lorentz gauge, as $A_\nu$ is proportional to the current
$j_\nu$, the gauge field components $A_1^a$ and $A_2^a$ along the
transverse direction are then small in magnitude in comparison
with those of $A_0^a$ and $A_3^a$ and can be neglected.  The
absence of the transverse gauge fields provides a needed
simplification for compactification.

The spatially one-dimensional string being an idealization of a more
realistic three-dimensional flux-tube, the description of produced
$q\bar q$ parton pairs within the string presumes the confinement of
these produced partons inside the string.  Hence, it is reasonable to
conceive further that the QCD4 compactification takes place under
transverse confinement.  We can describe transverse confinement in
terms of a confining scalar interaction $S({\bb r}_\perp)$ and the
quark mass function is then $m(\bb r_\perp)=m_0+S({\bb r}_\perp)$,
where $m_0$ is the quark rest mass.

Therefore, under the assumption of longitudinal dominance and
transverse confinement, the SU(N) gauge invariant action integral
in (3+1) Minkowski space-time is given by \cite{Pes95}:
\begin{eqnarray}\label{A4D}
 {\cal A} (4D) =  \int d^4 x && \Bigg\{ Tr  \left[ {\bar \Psi}
(4D, x)  \left( \gamma^\mu (4D)  (  i \partial_\mu +  g (4D) ~ T_a
A^a_\mu (4D, x) ) \  - ~ m (\bb r_\perp )\right)~ \Psi (4D, x) \right]
\nonumber
\\
&& \hspace*{1.0cm} - {1\over 16 \pi } F^a_{\mu \nu} (4D, x) \ F_a^{\mu \nu} (4D,
x) \Bigg\}, \label{1}
\end{eqnarray}
\begin{eqnarray}\label{eq2}
   F^a_{\mu \nu} (4D, x) &=&
\partial_\mu A^a_\nu (4D, x) - \partial_\nu A^a_\mu (4D, x) + g(4D) ~ f^a_{bc } ~ A^b_\mu (4D, x)~
A^c_\nu (4D, x)  ,
\end{eqnarray}
where $A^a_\mu (4D, x)$ and $\Psi (4D, x)$ are gauge and fermion
fields respectively with coordinates $x\equiv x^\mu = (x^0 , {\bb
  x})=(x^0 , x^1, x^2, x^3)$ and transverse coordinates $\bb
r_\perp=(x^1, x^2)$, $g=g(4D)$ is the coupling constant,
$\gamma^{\nu}(4D)$ are Dirac matrices, and $T_a$ are the generators of
the SU(N) group.

\subsection {Fermion part of the action integral}

We first examine ${\cal A } (4D, F) $, the part of the $4D$ action
integral in Eq.\ (\ref{A4D}) that involves the fermion field.  To carry out the
compactification, we write the Dirac fermion field $\Psi (4D , x )$ in
terms of functions $G_\pm({\vec r}_\perp)$ and
$f_\pm(x^0,x^3)$ \cite{Won91}:
\begin{eqnarray}\label{eq9}
&& \Psi (4D , x ) = \frac {1}{\sqrt{2}}\left( \begin{array}{cccc}
G_1  ({\bb r}_\perp ) \left( f_+ (x^0 ; x^3 ) + f_- (x^0 ; x^3 )
\right) \
  \\
 - G_2  ({\bb r}_\perp) \left( f_+ (x^0 ; x^3 ) - f_- (x^0 ; x^3 )
\right)
 \\
 G_1  ({\bb r}_\perp) \left( f_+ (x^0 ; x^3 ) - f_- (x^0 ; x^3 )
\right)
 \\
  G_2  ({\bb r}_\perp) \left( f_+ (x^0 ; x^3 ) + f_- (x^0 ; x^3 )
\right)
  \end{array} \right) .
\end{eqnarray}
 Using this explicit form of the Dirac fermion field $\Psi(4D, x)$, we
 carry out the simplifications and integrations over $x^1$ and $x^2$
 that eventually lead from ${\cal A } (4D, F)$ to ${\cal A } (2D, F)$,
\begin{eqnarray}
 {\cal A } (2D, F)  =  Tr  \int d^2 X
{\bar \Psi} (2D, X)\left[ \left( i \gamma^\mu (2D) \partial_\mu +
g (2D) \gamma^\mu T_a A^a_\mu (2D, X)\right)  -  m_{qT} \right]
 \Psi (2D, X)  ,
\label{10}
\end{eqnarray}
where $\mu=0,3$, and we have introduced the Dirac fermion field
$\Psi(2D)$, $\gamma$-matrices, and metric tensor $g_{\mu \nu}$,
according to the following specifications in (1+1)-dimensional
space-time in QCD2,
\begin{eqnarray}
&& \Psi ( 2D,X )= \left( \begin{array}{cccc} f_+ ( X ) \\
f_- ( X )
  \end{array} \right) , \ \ \ \ \ \ \ X = ( x^0 ; x^3 ) ,
  \label{11a}\\
  &&
\gamma^0 (2D)= \left( \begin{array}{cccc} 1 \ \ \ \ \ \ 0 \\ \\
0
    \ \ \ \ \ \ -1 \\
 \end{array} \right) , \ \ \  \gamma^3 (2D)   = \left( \begin{array}{cccc} 0 \
\ \ \ \ \ 1 \\ \\ -1 \ \ \ \ \ \ 0
 \end{array} \right), ~~~~~ g_{\mu \nu }(2D)= \left( \begin{array}{cccc} 1 \ \ \ \ \ \ 0 \\ \\
0
    \ \ \ \ \ \ -1 \\
 \end{array} \right).
\label{11}
\end{eqnarray}
The  $2D$ coupling constant $g(2D)$ is related to $4D$ coupling constant
$g(4D)$ by
\begin{eqnarray}
g(2D)=\int dx^1 dx^2 g(4D) [{ |G_1({\bb r}_\perp)|^2+|G_2({\bb
    r}_\perp)|^2}]^{3/2},
\label{13}
\end{eqnarray}
where the transverse wave functions $G_{1,2}({\bb r}_\perp)$ are
normalized according to
\begin{eqnarray}
\label{norm}   \int d x^1 d x^2 \left( \vert G_1 ({\bb r}_\perp)
\vert^2 +
  \vert G_2 ({\bb r}_\perp) \vert^2 \right) = 1 .
\label{14}
\end{eqnarray}
The transverse quark mass  $m_{qT}$ in Eq.\ (\ref{10}) is given
by
\begin{eqnarray}
 m_{qT} =   \int d x^1 d x^2  \left\{
m ({\bb r}_\perp)   \left( \vert G_1
  ({\bb r}_\perp) \vert^2 - \vert G_2 ({\bb r}_\perp) \vert^2
  \right) + \left[ \left( G^\ast_1 ({\bb r}_\perp) ( p_1 - ip_2 ) G_2 ({\bb  r}_\bot ) \right)
  - h.c.
   \right] \right\}.
\label{mqT}
\end{eqnarray}
In obtaining these results, we have considered $2D$ gauge fields
$A_\mu^a(2D,x^0,x^3)$ related to the $4D$-field gauge fields
$A_\mu^a(4D,x^0,x^3,{\bb r}_\perp)$ by
\begin{eqnarray}
\label{AA} A_\mu^a(4D,x^0,x^3,{\bb r}_\perp)= \sqrt{ |G_1({\bb
r}_\perp)|^2+|G_2({\bb
    r}_\perp)|^2} A_\mu^a(2D,x^0,x^3),~~~\mu=0,3.
\label{20}
\end{eqnarray}
The above equation means that along with the confinement of the
fermions, for which the wave function $G_{1,2}({\bb r}_\perp)$ is
confined within a finite region of transverse coordinates ${\bb
  r}_{\perp}$, we also consider the confinement of the gauge field
$A_\mu^a (4D , x)$, $\mu=0,3$, within the same finite region of
transverse coordinates, as in the case for a flux tube.

The result of Eq.\ (\ref{13}) reveals that as a result of the
compactification of QCD4, the coupling constant $g(2D)$ in lower
dimensional space in QCD2 acquires the dimension of a mass, and is
related to the confining wave functions of the fermions.  Fermions in
different excited states inside the tube will have different coupling
constants as indicated in Eq.\ (\ref{13}). The effective quark mass
$m_{qT}$ also depends on the transverse fermion wave functions, as
indicated in Eq.\ (\ref{mqT}).  In the lower two-dimensional
space-time, fermions in excited transverse states have a quark
mass different from those in the ground transverse states.

\subsection{Gauge field part of the action integral}

To go from ${\cal A}(4D,F)$ to ${\cal A}(2D,F)$ we have assumed
that the currents in the $x^0$ and $x^3$ directions are much
larger in magnitude than the currents in the transverse directions
so that $A_1^a$ and $A_2^a$ are small in comparison and can be
neglected. As a consequence, $F_{12}(4D)=0$ (we omit the
superscript symbol $a$ (color) for simplicity).  The evaluation of
all other components of $F_{\mu \nu}$ give for the gauge field
part of the action integral
\begin{eqnarray}
 \int \frac{d^4 x}{ 16 \pi }&&
 F^a_{\mu \nu}(4D) \ F_a^{\mu
\nu}(4D) = \int \frac{dx^0 dx^3}{ 16 \pi } \int dx^1 dx^2 (
|G_1({\bb r}_\perp)|^2+|G_2({\bb
  r}_\perp)|^2)  F^a_{03 } (2D)
 F_a^{03}(2D) \nonumber\\
&& - \int \frac{dx^0 dx^3}{ 16 \pi} \int dx^1 dx^2 \biggl (
\{\partial_1[{ |G_1({\bb r}_\perp)|^2
   +|G_2({\bb
    r}_\perp)|^2}]^{1/2} \}^2
~ +\{\partial_2[{ |G_1({\bb r}_\perp)|^2+|G_2({\bb
    r}_\perp)|^2}]^{1/2} \}^2\biggr )
\nonumber\\
&&\hspace*{3.0cm}\times [A_0(2D,x^0,x^3)A^0(2D,x^0,x^3)
+A_3(2D,x^0,x^3)A^3(2D,x^0,x^3)]. \label{28a}
\end{eqnarray}
It is useful to introduce the gluon mass $m_{gT}$ that arises from
the confinement of the gluons in the transverse direction,
\begin{eqnarray}\label{eq31}
 m_{gT}^2= {1\over 2} \int dx^1 dx^2
\left[  \Big\{\partial_1 \left({ \sum\limits_{i=1}^2 \vert G_i
({\vec r}_\bot ) \vert^2}\right)^{1/2} \Big\}^2+ \Big\{\partial_2
\left({ \sum\limits_{i=1}^2 \vert G_i ({\vec r}_\bot )
\vert^2}\right)^{1/2} \Big\}^2 \right].
\end{eqnarray}
As this gluon mass $m_{gT}$ arises from the confinement
compactification of the gluon within the flux tube, we can call
such a mass the compactification
 mass of the gluon.
Equation (\ref{28a}) becomes
\begin{eqnarray}\label{eq32}
 \int \frac{d^4 x}{ 16 \pi }
 F^a_{\mu \nu}(4D) \ F_a^{\mu
\nu}(4D) &=&  \int \frac{dx^0 dx^3} {16 \pi }
 \biggl \{ F^a_{03 } (2D) F_a^{03}(2D) - 2 m_{gT}^2  [A_\mu^a(2D)A^\mu_a(2D)
] \biggr \}
\end{eqnarray}
We collect all the fermion and gauge field parts of the action
integral in ${\cal A} (4D )$ in Eq.\ (\ref{1}).  All terms in the
action integral ${\cal A} ( 4 D )$ (including matrices and
coefficients) are in the Minkowski $(1+1) $ dimensional
space-time. We can rename the action integral ${\cal A} (4D )$ to
be ${\cal A} (2D )$ given explicitly by
\begin{eqnarray}
{\cal A} (2D ) = \int d^2 X && \Bigg\{  Tr \Bigg[  {\bar \Psi}
    (2D,X) \bigg[ \gamma^k(2D) \left(     i \partial_\mu +  g (2D) ~ T_a A^a_\mu (2D,
x)    \right) - m_{qT} \bigg]\Psi (2D,X)\nonumber \\
&&\hspace*{1.0cm}
  -{1\over 16 \pi } F^a_{\mu \nu} (2D)
  \ F_a^{\mu \nu} (2D)  +{1\over 8 \pi }m_{gT}^2 [A_a^\mu(2D)A^a_\mu(2D)]   \Bigg\}. \label{31a}
\end{eqnarray}
Thus, in the presence of longitudinal dominance and transverse
confinement, we succeed in compactifying the action integral from
${\cal A} ( 4 D )$ in QCD4 to ${\cal A} ( 2 D )$ in QCD2, by
introducing $g_{2D}$, $m_{qT}$, and $m_{gT}$ that contain information
about the transverse profile. All the transverse flux tube information
is subsumed under these quantities.  In this way, the $2D$ gauge field
appears to be massive where $m_{gT}$ and $m_{gT}$ arise as a
consequence of the transverse confining motion of both fermion and
gauge fields. The physical explanation of such effect consists in the
decrease of the number of trajectories in moving from one point of the
space to another point, as a direct consequence of
compactification. Such constraints in movement manifest themselves as
masses of field particles. The magnitudes of $m_{qT}$ and $m_{gT}$
depend strongly on the kind of the compactification that is dictated
by the transverse functions $G_1({\bb r}_\perp)$ and $G_2({\bb
  r}_\perp)$ (see Eqs.(\ref{mqT}) and (\ref{eq31})).

\section{Equations of a transverse motion in a tube
and Fermion effective mass}

To obtain the equations of motion for  the functions $G_1  ({\vec
r}_\bot )$ and $G_2  ({\vec r}_\bot )$,  we vary the action
integral ${\cal A} (4D)$ (\ref{1}) with the fermion fields $\Psi
(4D , x)$ given by Eq.(9), under the constraint of the
normalization condition, (\ref{norm}). To do this we construct a
new functional ${\cal F}$
\begin{eqnarray}\label{eq80}
&& {\cal F} = {\cal A }(4D ) + {\lambda\over 2}  \int d x^1 d x^2
\left( \sum\limits_{i=1}^2 \vert G_i ({\vec r}_\bot ) \vert^2
\right) \int d x^0 d x^3 \left( {\bar \psi} (2D,X) m_{qT}  {\psi}
(2D,X)  \right),
\end{eqnarray}
where $\lambda$ is the Lagrange  multiplier. The last term in Eq.\
(\ref{eq80}) takes into account the unitarity of a fermion field
in the 4D space-time.  Varying the last equation with respect to
the functions $G_1 ({\vec r}_\bot )$ and $G_2  ({\vec r}_\bot )$
we derive
\begin{eqnarray}\label{eq81}
&& ( p_1 +  ip_2 )  G_1 ({\vec r}_\bot )  = (  m ({\vec r}_\bot )
+\lambda )  G_2  ({\vec r}_\bot ), ~~~~~~  ( p_1 - ip_2 ) G_2
({\vec r}_\bot ) = (\lambda   - m  ({\vec r}_\bot ) )
G_1  ({\vec r}) ,\nonumber \\
&&( p_1 +  ip_2 )  G^\ast_2 ({\vec r}_\bot )  = (  m ({\vec
r}_\bot ) - \lambda )  G^\ast_1  ({\vec r}_\bot ), ~~~~~ ( p_1 -
ip_2 ) G^\ast_1 ({\vec r}_\bot ) = - (  m ({\vec r}_\bot ) +
\lambda ) G^\ast_2 ({\vec r}).
\end{eqnarray}
Carrying out complex conjugation in the   last   two equations we
obtain $\lambda = \lambda^\ast$. Substituting the equations
(\ref{eq81}) for $G_{1,2} ({\vec r}_\bot )$ functions into the
formula (\ref{mqT}) for $m_{qT}$ we get
\begin{eqnarray}\label{eq84}
 m_{qT}& = & \lambda .
\end{eqnarray}
Thus, the effective mass of the compactified 2D fermion field is equal
to the energy of the transverse motion of the 4D fermion, which is an
eigenvalue of Eq.\ (\ref{eq81}). We should note here that the 2D
fermion can generally gain a mass even when the initial 4D fermion
appears to be massless. The explanation of such phenomenon is the same
as before. The compactification effectively leads to constraints in
moving a fermion from one point of a space-time to another one due to
the decrease of the number of trajectories in the 2D space-time, as
compared to the 4D space-time.

\section {Conclusion}

 Under the assumption of longitudinal dominance and transverse
 confinement, the SU(N) gauge invariant field theory of QCD4 can be
 compactified into QCD2 in Minkowski $(1+1)$ dimensional space-time
 from the consideration of the action integral. The compactified 2D
 action integral ${\cal A} (2D)$ depends only on 2D-fields.  The
 corresponding coupling constants, effective quark, and gluon masses
 in two-dimensional space-time are derived.  Such compactification
 leads to strong changes in physics of the 2D Lagrangian that is
 manifested in both the renormalization of coupling constant and
 fermions as well as gauge field bosons acquiring masses.

 \vspace*{0.3cm}

\centerline{\bf Acknowledgment}

\vspace*{0.3cm}   The research  was supported in part by the
Division of Nuclear Physics, U.S. Department of Energy.

\end{document}